# Polarisation-driven magneto-optical and nonlinear-optical behaviour of a room-temperature ferroelectric nematic phase

Evangelia Zavvou
*Otto von Guericke University Magdeburg, Institute of Physics,
Dept. Nonlinear Phenomena, Magdeburg, Germany
and
Department of Physics,
University of Patras, 26504, Patras, Greece*

Melanie Klasen-Memmer, Atsutaka Manabe, and Matthias Bremer
*Electronics Division, Merck KGaA, Darmstadt, Germany*

Alexey Eremin*
*Otto von Guericke University Magdeburg, Institute of Physics,
Dept. Nonlinear Phenomena, Magdeburg, Germany*

Nematics with a broken polar symmetry is one of the fascinating recent discoveries in the field of soft matter. High spontaneous polarisation and the fluidity of the ferroelectric nematic $N_\text{F}$ phase make such materials attractive for future applications and interesting for fundamental research. Here we explore the polar and mechanical properties of a room-temperature ferroelectric nematic and its behaviour in a magnetic field. We show that $N_\text{F}$ is much less susceptible to the splay deformation than to the twist. The strong splay rigidity can be attributed to the electrostatic self-interaction of the polarisation avoiding the polarisation splay.

## I. INTRODUCTION

Nematics are the simplest examples of partially ordered soft condensed matter systems with only the long-range quadrupolar molecular orientational orders [1, 2]. Molecular chirality extends the family of the nematics to cholesterics and blue phases, which additionally exhibit helical structure [3–5]. Nematics with other, more "exotic" kinds of symmetries, such as biaxial, polar, tetrahedratic, etc., have been predicted in the frame of mean-field theory for bend-shaped mesogens [6–9].

Among those phases, recently discovered nematics with polar symmetry stand out. Such ferroelectric/ferroelastic nematics are distinguished for their remarkable properties such as large electric susceptibility, pyroelectric, and second-order nonlinear optics coefficients [10–17]. However, the mechanisms of stabilising the polar order are still under intensive research [18–22]. A large dipole moment, wedge-shape of the mesogens, and dense packing are prerequisites for the ferroelectric order, resulting in breaking the molecular head-tail symmetry [15].

One of the consequences of the ferroelectric nematic order is the instability towards the splay state predicted theoretically [8, 20, 23]. Instability of the polar nematic towards the twist deformation was studied by Khacheturyan [24]. High electric polarisation, in the order of μC cm$^{-2}$ makes those materials promising for low-power, fast-switching electro-optical applications and energy storage [25].

Fréedericksz transition underlying realignment of the director in nematics is in the heart of many LC applications [2]. Dielectric or diamagnetic torques resulting from the anisotropy of the nematics drive the director's reorientation. The anchoring at the substrates impedes the realignment, restricting the orientation of the mesogens at the boundaries. Consequently, a continuous deformation of the director is obtained across the cell. The ground state of a ferroelectric nematic under given electrical conditions is determined by the minimum of the free energy, which must include electrostatic contributions. However, the electrostatic interactions arising from the bound charges and the spontaneous polarisation complicate the description. For instance, spatially inhomogeneous charges can develop and be localised near the electrodes. Generally, the free energy of the ferroelectrics becomes shape-dependent.

In a ferroelectric nematic with a high spontaneous polarisation, the splay director deformation results in the polarisation splay disfavouring the director reorientation. In addition to the diamagnetic torque, electrostatic contribution is also important during the reorientation in a magnetic field. Observation of a magnetic Fréedericksz transition is perhaps the simplest case, where the deformation is driven by the magnetic field and counteracted by the elastic torques. The electrostatic field developing inside the nematic is solely determined by the inhomogeneity of the polarisation.

---

* alexey.eremin@ovgu.de

The situation becomes more complicated in an electric field due to the direct coupling between the polarisation, its splay, and an applied electric field [26]. In this paper, we explore the behaviour of a room-temperature ferroelectric nematic compound [17] in magnetic and electric fields using optical transmission measurements, second harmonic generation, dielectric spectroscopy and confocal laser scanning microscopy. We develop a model describing the director reorientation in a magnetic field and compare its predictions with the experimental observations. We demonstrate that the electrostatics can have major effect on the mechanical properties of the ferroelectric nematic.

## II. EXPERIMENTAL

The investigated mesogen is 4-((4'-butyl-2',3,5,6'-tetrafluoro-[1,1'-biphenyl]-4-yl) difluoromethoxy)-2,6-difluorobenzonitrile exhibiting a monotropic ferroelectric nematic phase near room temperature. The synthesis and the initial characterisation are published in [17]. The net dipole moment of the molecule is 11.3 D and the static dielectric permittivity $\varepsilon'$ in $N_F$ reaches a remarkable value of 20000.

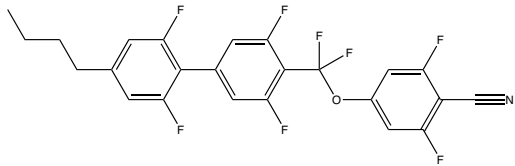

Cr (-7°C $N_F$) 19.6°C iso

FIG. 1: Chemical structure of the investigated mesogen

Polarisation microscopy studies were made using AxioImager A.1 polarising microscope (Carl Zeiss) equipped with a heat-stage (Instec, USA). The samples were prepared in commercial glass cells (Poland) equipped with planar transparent indium tin oxide (ITO) electrodes (cell thickness: 30 µm, ITO resistance: 10 Ω), and rubbed polyimide layers for LC planar alignment. Cells with interdigitated in-plane electrodes (IPS) were used for studying the in-plane switching.

Birefringence measurements were carried out in an Axioskop 40 pol polarising microscope (Carl Zeiss) equipped with a Berek tilting compensator (Leitz) and a Linkam LTS420 hotstage with a LNP96 cooling pump. Measurements were acquired under monochromatic light of wavelength $\lambda = 546$ nm. The material was filled to commercial cells for planar alignment with 5 µm thickness (Instec, USA).

Fréedericksz transition was studied through measurement of the optical transmission between two crossed polarising prisms (Thorlabs). For the magnetic Fréedericksz transition, the cell was placed between two Helmholtz coils with the magnetic field along the cell normal. The magnetic field was generated by a pair of electromagnets with the maximal magnetic flux density of 650 mT. Measurements of the optical transmission were performed upon increasing magnetic field up to 650 mT with a step of 10 mT and 15 s waiting time between the field application and the data acquisition. Critical field for the magnetic Fréedericksz transition was also determined through dielectric spectroscopy. Dielectric measurements were performed employing an Alpha-N frequency response analyser (Novocontrol). Isothermal curves of the dielectric permittivity as function of frequency were acquired between 0.1 Hz and 1 MHz using an alternating field of $V_{\text{rms}} = 0.1$ V. Succesively increasing values of the magnetic field were applied, varying from zero up to 1.4 T in steps of roughly 60 mT. For the evaluation of the critical magnetic field for Fréedericksz transition the frequency of $f = 110$ Hz was used.

Polarizing confocal laser scanning microscopy (CLSM) studies were conducted using a Leica TCS SP8 CLSM microscope on samples doped with 0.01 wt% of a dichroic dye (N,N'-bis(2, 5-di-tertbutylphenyl) - 3,4,9,10 - perylenedicarboximide (BTBP), Sigma-Aldrich) excited at $\lambda = 488$ nm. Generation of the optical second harmonic (SHG) was measured using the multiphoton laser of the TCS SP8-Leica microscope. A tunable IR laser ($\lambda = 880$ nm) was used as a fundamental light beam.

## III. RESULTS

### A. Polarisation switching

Upon cooling of the isotropic liquid below $T = 20.6$ °C, nematic phase nucleates in form of birefringent droplets, as can be seen in Fig. 2a, which coalesce into a nearly uniform texture in an LC cell with planar alignment Fig. 2b. The optical texture exhibits typical for nematics director fluctuations, and the birefringence only weakly depends on the temperature (Fig. 3).

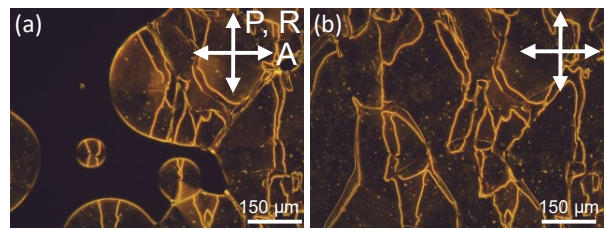

FIG. 2: Optical microscopy textures of the studied compound taken between crossed polarisers in a 30 µm planar cell at: (a) 20.2 °C and (b) 18 °C.

Bistable switching in an electric field is one of the essential features of ferroelectrics, which distinguishes them from other polar materials [27]. Polarisation reversal drives the electric switching in liquid crystals, which is often accompanied by the reorientation of the optical axes and/or birefringence change (optical switching). The polarisation reversal in a ferroelectric sample confined in a capacitor cell results in the current transient, which is indicative of the ferro- or antiferroelectric character of the switching [28]. Figure 4a shows the current transients recorded in the $N_F$ phase at $T = 13\,°C$ at various amplitudes of triangular-wave voltage applied to the test cell. At small amplitudes, we observe a single current peak centred at the voltage $V_{th} = 0.16\,V\,\mu m^{-1}$. As the amplitude increases, a second broad peak appears at high fields. Multiple peaks in the current transients have been observed in various LC systems[29–33]. They can be attributed to the antiferro- or ferrielectric types of order, complex anchoring conditions, trapping of charges in the topological defects and dislocations, and the ionic impurities. The ionic contribution becomes dominant at high fields resulting in a broad current peak.

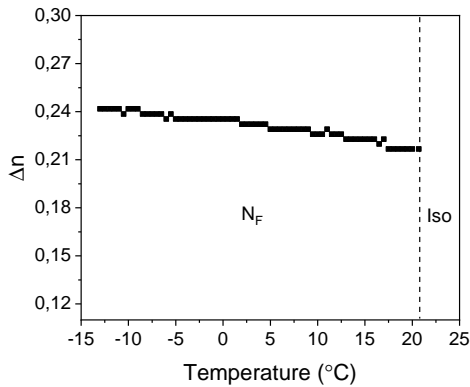

FIG. 3: Temperature dependence of the birefringence of the investigated mesogen measured in a 5 µm planar cell.

The switching polarisation $P_{sw}$ was estimated from the area under the current reversal peak, which is plotted in Figure 4b. $P_{sw}$ is very high, reaching almost $6\,\mu C\,cm^{-2}$. Switching polarisation can be considered as an estimation of the spontaneous polarisation of the liquid crystal [28]. A discontinuous increase of $P_{sw}$ upon the isotropic- $N_F$ transition reflects the first-order character of the transition. The polarisation only slightly increases on decreasing temperature. Below $T = 11.5\,°C$, however, there is a continuous decrease in the polarisation which is driven by slow crystallisation kinetics of the monotropic $N_F$ phase.

### B. Nonlinear optical response

Generation of the optical second harmonic (SH) generation (SHG) is a widely used technique to establish the polar order in crystals and liquid crystals [34–36]. In contrast to the switching current measurements, it does not require poling of the polarisation and application of external fields. SHG results from the nonlinear interac-tion between the light and the media, which leads to the energy transfer from the primary beam to the SH beam. The phase symmetry determines the allowed components of the second order optical susceptibility tensor and forbids SHG in the phases with the inversion symmetry.

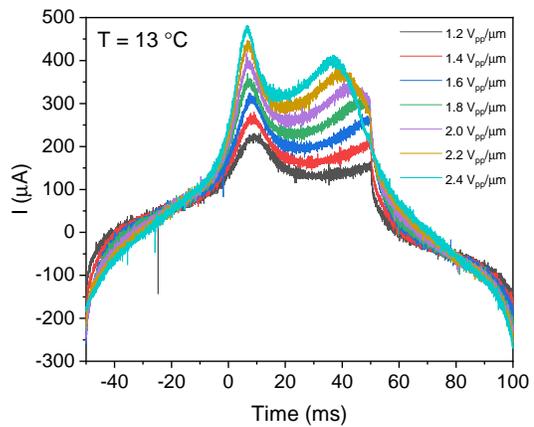

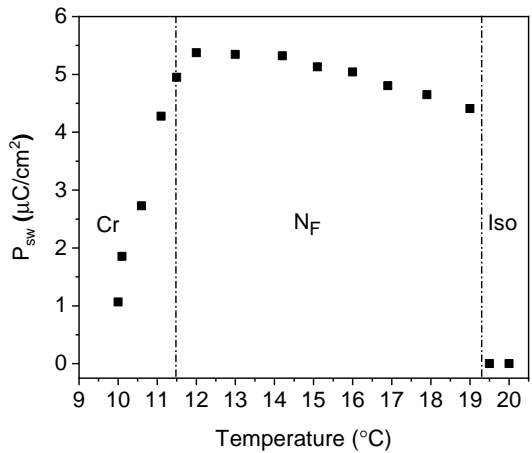

FIG. 4: Polarisation switching in a planar cell 30 µm cell: (a) current transients recorded at $T = 13\,°C$ at various amplitude of the triangular-wave field. (b) Temperature dependence of the switching polarisation $P_{sw}$ measured at a frequency $f = 5$ Hz and amplitude $0.7\,V\,\mu m^{-1}$

To study spatially-resolved generation of the SH, we employ SHG microscopy technique, where an IR laser (880 nm) is used as the light source in the confocal scanning microscope. The microscopic textures in untreated glass slides are shown in Figure 5a between crossed polarisers and in Figure 5b the corresponding SHG texture.

To distinguish SHG from the multiphoton fluorescence with a broad spectrum, we recorded the transmission spectrum of the sample under 880 nm excitation (Fig. 5c). The spectrum features only a single resolved peak centred at 440 nm, suggesting that our sample is SHG active. An extraordinary large SHG signal occurs spontaneously, without any applied external field, confirming the polarity of the ground state of the nematic phase.

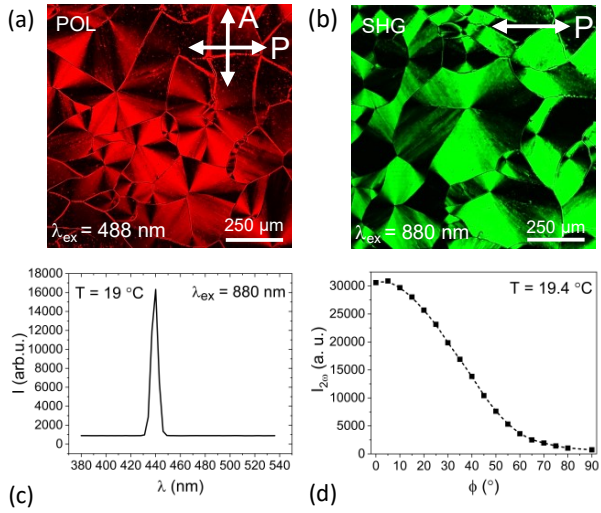

FIG. 5: (a) Polarisation microscopy image taken at 18.9 °C between a pair of untreated glass plates (b) SHG microscopy image of the texture in (a); (c) the spectrum of the SHG signal taken with excitation at $\lambda_{ex}$ =880 nm; (d) the angular dependence of the SHG signal in a homogeneously aligned sample.

Angular dependence of the of the SHG polarisation in Figure 5c established the relation between the polar and the nematic directors. The highest intensity of the SH signal corresponds to the polarisation along the nematic director. The temperature dependence of the SHG signal is shown in Figure 6. As the temperature decreases, the signal increases by several orders of magnitude.

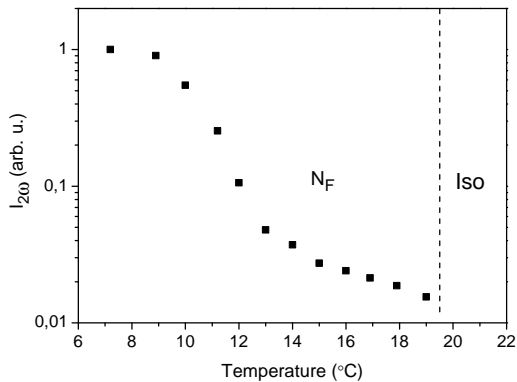

FIG. 6: Temperature dependence of SHG signal measured in untreated glass slides.

The spontaneous polarity can be very well demonstrated in by confocal SHG microscopy images of sessile droplets on a glass substrate (Fig 7). Between crossed polarisers, the optical image features an extinction cross with a disclination line in the centre. The director orientation can be determined from the angular dependence of the absorption by the dichroic dye dissolved in the liquid crystal (Figure 7b,d). The intensity distribution suggest the tangential alignment of the director arranged into concentric rings. The intensity distribution confirms our assumption that the polarisation is parallel to the nematic director.

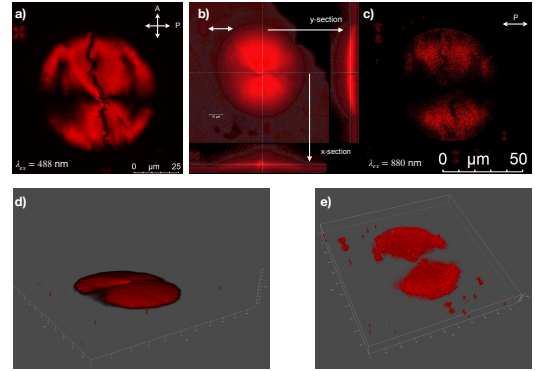

FIG. 7: Sessile droples imaged by the confocal microscopy: a) Transmission POM image taken in a monochromatic light, b) three orthogonal section of the confocal microscopy image (fluorescence), c) a horisontal section of the confocal SHG image, d) 3D reconstruction of the sessile droplet in b) as observed in the confocal polarising microscopy, e) 3D reconstruction of the confocal SHG image of the droplet in c).

Bistable character of the polarisation switching is evident from the field-dependent SHG measurements in the cells with in-plane electrodes (IPS). The signal $I^{2\omega}(E)$ exhibits a distinct hysteresis behaviour (Fig. 8). A coercive field of the opposite polarity is required to suppress the spontaneous signal.

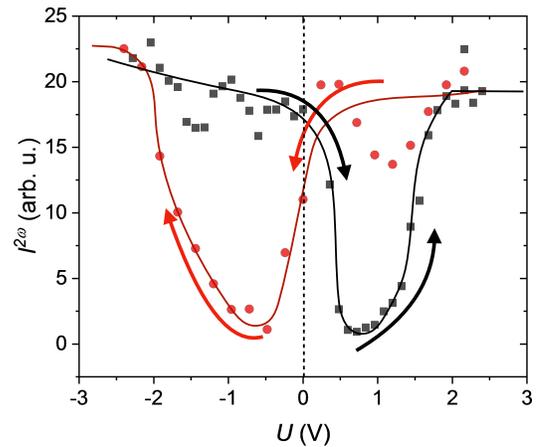

FIG. 8: Hysteresis-type field-dependence of the SHG signal $I^{2\omega}$ recorded at $T = 15\,°C$ (the lines are guide for the eye)

## C. Fréedericksz transition

In contrast to the electric Fréedericksz transition (FT), the magnetic FT allows a much more straightforward interpretation since the magnetic field does not directly couple to the electric polarisation or charges. The critical magnetic field is determined by the diamagnetic anisotropy $\chi_a$ and the Frank elastic constant $K$. Since this compound does not exhibit a regular nematic phase, we cannot estimate $\chi_a = \chi_\parallel - \chi_\perp$ from the simultaneous measurements of the critical magnetic and electric fields. Instead, we use the increment system developed by Flygare et al. [37] to calculate $\chi_a$ from the chemical structure of the mesogen. Aromatic moieties give the largest contribution to diamagnetic susceptibility. We assume that the main axis of the mesogen is given by the tetrafluoro-biphenyl part, and the difluorobenzonitrile head group can freely rotate around the main axis at an angle of 65.3 °. From those assumptions, we estimate $\chi_a = 1.607 \times 10^{-6}$. The Fréedericksz transition was studied using the measurements of the optical transmission, and the critical field $\mu_0 H_c$ was determined from the extrapolation of the $I(H)$ curves. In the splay geometry, the critical field is about 580 mT in a 30-micrometre thick cell, which is close to the limit of our measuring magnet. The critical field in the splay geometry was also investigated through dielectric spectroscopy measurements in a 30 µm planar cell. The acquired isothermal curves of the real part of dielectric permittivity versus frequency are presented in Figure 9a, for increasing values of applied magnetic field. Although the compound exhibits very high values of dielectric permittivity, this is not the case for dielectric anisotropy, which is two orders of magnitude smaller. It should be noted, that within the $N_F$ phase, values of dielectric permittivity measured in cells with polyimide aligning layers are effectively smaller due to the capacitance of the aligning layers [38]. However, this phenomenon does not affect dielectric anisotropy and the determination of the critical field of splay FT. The detailed investigation of the dielectric response of the compound is out of the scope of the present work. The critical magnetic field of splay Fréedericksz transition was determined at $f = 110$ Hz, away from any relaxation mechanism, and at 18 °C it was estimated around 560 mT, as shown in Figure 9b, being in agreement with optical transmission measurements.

The values of the critical magnetic field determined from optical transmission measurements correspond to a splay elastic constant $K_{11}$ as high as 36 pN. $K_{11}$ shows hardly any temperature dependence, as shown in Figure 10. At the same time, the twist elastic constant $K_{22}$ appears significantly smaller and amounts only 6.4 pN.

The total free energy of the cell is the sum of the Frank-Oseen energy of the director deformation $F_{FO}$, magnetic energy $F_m$, and the electrostatic energy $F_{es}$ induced by the director deformation inside the cell:

$$F = F_{FO} + F_m + F_{es} \tag{1}$$

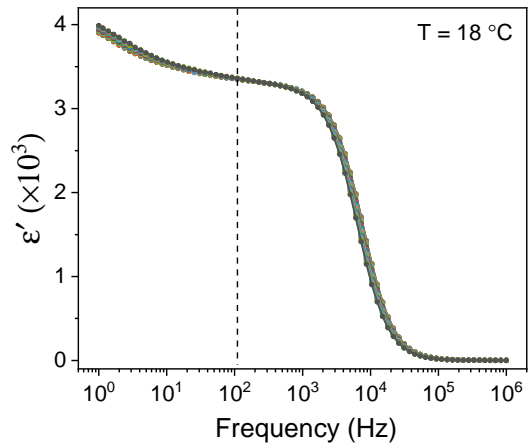

(a)

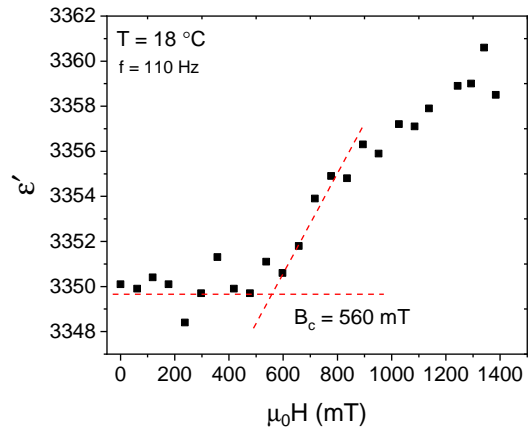

(b)

FIG. 9: (a) Isothermal curves of the dielectric permittivity as a function of frequency at 18 °C measured in a 30 µm cell with polyimide rubbed substrates.(b) Magnetic field dependence of the real part of the dielectric permittivity at $f = 110$ Hz
.

However one has to consider the polar structure of the nematic phase. In the theory of ferroelectrics the polarisation contribution to the field-independent part of Landau free energy is given by [39]

$$f_p = \frac{1}{2} A \mathbf{P}^2 + \frac{1}{6} \mathbf{P}^4 + \frac{\gamma}{2} (\nabla \mathbf{P})^2 \tag{2}$$

Landau-type theory of polar nematics was developed by Brand and Pleiner [40]. The authors considered the polarisation as a primary order parameter and discussed the stability of the phase with respect to the formation of the splay phase. The free energy expansion contains the splay, bend and spontaneous splay terms in addition to the expansion on powers of $P$.

In the case of $N_F$, however, $\mathbf{P}$ and the orientational Q-tensor can be considered as two coupled order parameters. The theories using this approach were developed for the twist-bend nematic phase by Selinger et al. [41], and by Kats [23] for $N_F$. These models contain the flexoelectric couplings between the director and the polarisation as well as the linear coupling. Those theories can predict both uniform and modulated phases.

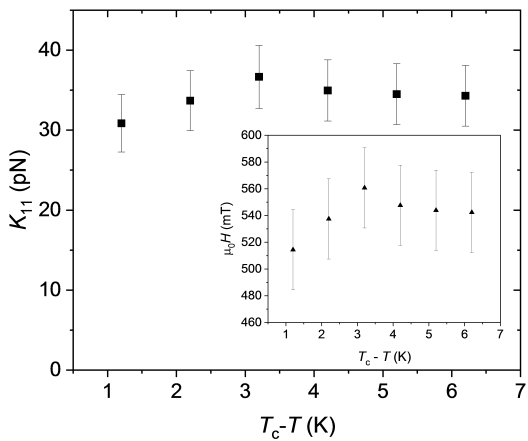

FIG. 10: Temperature dependence of the splay elastic constant measured in a 25 μm cell with polyimide rubbed substrates. The critical magnetic field $\mu_0 H_c$ is given in the inset.

Here we consider the nematic state with the established uniform polar order along the director $\mathbf{n}$ with the spontaneous polarisation $\mathbf{P}_s = P_s \mathbf{n}$. The geometry is defined by a plane parallel cell with the planar anchoring of the director $\mathbf{n}$ at the interfaces and uniform alignment in the bulk. Assuming the polar character of the director, the nematic adopts a uniform spontaneous polarisation $P_s$ parallel to the substrate. A magnetic field is applied in the direction perpendicular to $\mathbf{n}$ exerting a magnetic torque on the director. In one-constant approximation, the expressions for the Frank-Oseen energy density is given by

$$f_{\rm FO} = \frac{K}{2}\left(\frac{{\rm d}\theta}{{\rm d}z}\right)^2 \tag{3}$$

and the magnetic energy density is

$$f_{\rm m} = -\frac{\chi_a B^2 \cos^2 \theta}{2\mu_0} \tag{4}$$

where $K$ is the effective Frank elastic constant, $B$ is the magnetic flux density, $\chi_a$ magnetic anisotropy, $\mu_0$ is the magnetic permittivity of vacuum.

The Fréedericksz transition in the splay geometry is accompanied by the formation of the director splay areas near the substrate. The polarisation becomes nonuniform exhibiting a dependence on the vertical coordinate $z$ with a vertical component $P_{s,z} = P_s \sin\theta(z)$ see Eq. 5, where $\theta(z)$ is the angle between the director $\mathbf{n}$ and the polarisation $\mathbf{P}$. Polarisation splay results in a nonuniform distribution of the bound charges

$$\rho_{\rm b} = -\nabla \cdot \mathbf{P} = P_s \cos\theta \cdot \theta' \tag{5}$$

where $\theta' = {\rm d}\theta/{\rm d}z$.

This charge distribution determines the electric potential $\varphi(z)$ $\rho_b(z) = -\nabla \cdot \mathbf{P}_s$ and can be obtained from the solution of the Poisson equation

$$\nabla^2 \varphi = \frac{\nabla \cdot \mathbf{P}}{\varepsilon_0 \varepsilon} \tag{6}$$

Substituting $P_{s,z} = P_s \sin\theta(z)$ into Eq. 6, we obtain $\varphi''(z) = (\varepsilon_0 \varepsilon)^{-1} P_s \cos\theta \cdot \theta'$ yielding the solution for the case of the constant charge (open circuit)

$$\varphi(z) = \varphi_0 + \frac{P_s d}{\varepsilon_0 \varepsilon} \int_1^{z/d} \sin\theta(t) {\rm d}t \tag{7}$$

The electrostatic contribution is given by the term

$$F_{\rm ES} = \int_0^d (\rho_b + \rho)\varphi {\rm d}V \tag{8}$$

where $\rho_b$ and $\rho$ are the densities of the bound and free charges, respectively.

Neglecting the free charges in the system and substituting the solution for the potential $\varphi(z)$, we obtain the energy density

$$f_{\rm ES}(\theta'(z), \theta(z), z) = (\rho_b + \rho)\varphi =$$
$$= -\frac{P_s^2 d}{\varepsilon_0 \varepsilon} \theta' \cos\theta \int_1^{z/d} \sin\theta(\xi){\rm d}\xi \tag{9}$$

The equilibrium equations can be obtained for arbitary dependence $\theta = \theta(z)$ form using variation principles. The net energy is given by the sum of the elastic, magnetic and electric contributions:

$$f = \frac{1}{2} K \theta'(z)^2 + \frac{\chi_a B^2 \cos^2\theta(z)}{2\mu_0} -$$
$$- \frac{P_s^2 d}{\varepsilon_0 \varepsilon} \theta' \cos\theta \int_1^{z/d} \sin\theta(\xi){\rm d}\xi \tag{10}$$

The equilibrium condition is described by

$$\delta \tilde{F} = 0 \tag{11}$$

where the variation $\delta \tilde{F}$ is given by

$$\delta \tilde{F} = K\theta''(z) + \frac{\chi_a B^2 \sin\theta(z)\cos\theta(z)}{\mu_0} -$$
$$- \frac{P_s^2 \sin\theta(x)\cos\theta(z)}{\epsilon_0 \epsilon} \tag{12}$$

This equation can be converted to a dimensionless form by introducing the characteristic field $B_{\rm c} = \frac{1}{d}\sqrt{\frac{\mu_0 K}{\chi_a}}$, and the dimensionless field $\beta = B/B_c$. The contribution of the spontaneous polarisation term is scaled by the parameter

$$\kappa_p^2 = \frac{P_s^2 d^2}{\pi^2 \varepsilon_0 \varepsilon K} \tag{13}$$

yielding

$$\theta'' + (\beta^2 - \kappa_p^2)\sin\theta\cos\theta = 0 \tag{14}$$

The Fréedericksz transition sets off when $\beta^2 - \kappa_p^2 = \pi^2$ or the external magnetic field satisfies

$$B_c^2 = \frac{\pi^2}{d^2}\frac{\mu_0}{\chi_a}K\left(1 + \kappa_p^2\right) \tag{15}$$

The elastic constant $K$ becomes effectively rescaled by the polarisation contribution. The characteristic parameter $\kappa_p$ determines the scale of the polarisation-splay effect. To estimate the value of $\kappa_p$, we take the values of the switching polarisation as $P_s = 6\,\mu\text{C}\,\text{cm}^{-2}$, typical value for the effective elastic constant $K = 10\,\text{pN}$, $\varepsilon = 1 \times 10^4$, $d = 30\,\mu\text{m}$, we estimate $\kappa_p = 10^7$. This is a very large value, which appears to contradict the experimental observations. One possibility to explain this contradiction is to assume that the estimations of the spontaneous polarisation are not accurate. The value $P_s = 6\,\mu\text{C}\,\text{cm}^{-2}$ is large even for some solid ferroelectrics [42]. Current-response measurements of the $P_s$ can be strongly biased by the induced polarisation due to the ionic contributions and the induced polarisation due to a very high dielectric permittivity [29, 30]. In smectic LCs, materials exhibiting a high switching polarisation become antiferroelectrics rather than ferroelectrics [27]. An electric field of a uniformly polarised dielectric slab cut perpendicular to **P** with such high polarisation is as high as $P/\varepsilon_0\varepsilon = 0.68\,\text{V}\,\mu\text{m}^{-1}$, which roughly corresponds to 20 V across a typical 30 μm cell. At the same time, the induced polarisation in typical for the experiment field of $E = 10\,\text{V}\,\mu\text{m}^{-1}$, is $P = \varepsilon_0(\varepsilon - 1)E \approx 80\,\mu\text{C}\,\text{cm}^{-2}$. As a result of screening, the polarisation switching is observed when the field is applied parallel to the thin LC slab. Electrostatic contribution to the stiffening of $K$ can be weakened by the partial screening of the electrostatic interactions by ionic impurities. The last possibility could be the softening of the splay constant driven by the modulation towards the spontaneous splay expected to occur in polar nematics.

## IV. CONCLUSIONS

We investigated the polar and magneto-optical behaviour of the room-temperature ferroelectric nematic phase with a high switching polarisation. The first order isotropic-$N_F$ transition is marked by a strong spontaneous nonlinear optical response (SHG). The polar axis is aligned along the nematic director. We show that the splay elastic constant exceeds the twist constant considerably with $K_{11}/K_{22} \approx 6$. In a comparable mesogen such as 5CB, for example, this ratio is about 2. Another important finding is the absolute value of $K_{11}$ above 30 pN for a compound having such a low isotropic-(Ferro)nematic transition temperature is very high. For comparison, $K_{11}$ is in the order of 6.4 pN at room temperature. High spontaneous polarisation can be responsible for the stiffening of the splay elastic constant since the polarisation splay increases the electrostatic energy of the system. At the same time, the twist deformation remains unaffected by the electrostatics.

## V. ACKNOWLEDGMENTS

The authors acknowledge Panagiota K. Karahaliou and Alexandros G. Vanakaras (University of Patras, Greece); Nerea Sebastián and Alenka Mertelj (Jožef Stefan Institute, Slovenia) for fruitful discussions and Deutsche Forschungsgemeinschaft (Project ER467/8-3) for the financial support.